\documentclass{vow2008}

\title{Virtual Observatory for Astronomers: Where Are We Now?}
\author{Igor V. Chilingarian}
\affil{Observatoire de Paris-Meudon, LERMA, UMR~8112, 61 Av. de l'Observatoire, Paris, 75014, France}
\affil{Sternberg Astronomical Institute, Moscow State University, 13 Universitetski prospect, 119992, Moscow, Russia}
\affil{VO-Paris Data Centre, Observatoire de Paris, France}

\usepackage{graphicx}
\begin{document}


\maketitle

\begin{abstract} 
After several years of intensive technological development Virtual
Observatory resources have reached a level of maturity sufficient for their
routine scientific exploitation. The Virtual Observatory is starting to be
used by astronomers in a transparent way. In this article I will review
several research projects making use of the VO at different levels of
importance. I will present two projects going further than data mining: (1)
studies of environmental effects on galaxy evolution,
where VO resources and services are used in connection with dedicated
observations using a large telescope and numerical simulations, and (2) a
study of optical and near-infrared colours of nearby galaxies complemented
by the spectroscopic data.
\end{abstract}

\section{Introduction}

What is the Virtual Observatory? The Virtual Observatory is a realisation of
the e-Science concept in astronomy; it is a \emph{powerful virtual
environment} aimed at facilitating astronomical research and increasing
scientific output of astronomical data. It is formed by data archives and
software tools interoperating using a set of peer-reviewed standards and
technologies developed and maintained by the International Virtual
Observatory Alliance (IVOA).

What does this really mean? Being naive, the increase of the scientific
output of the data means that each Gigabyte of data coming
from a given instrument will produce larger number of scientific results,
i.e. papers or conference presentations, exactly like uploading a research
paper to the preprint server increases significantly its scientific impact.

\section{Similarities between VO and WWW}

\begin{figure}
\includegraphics[width=\hsize]{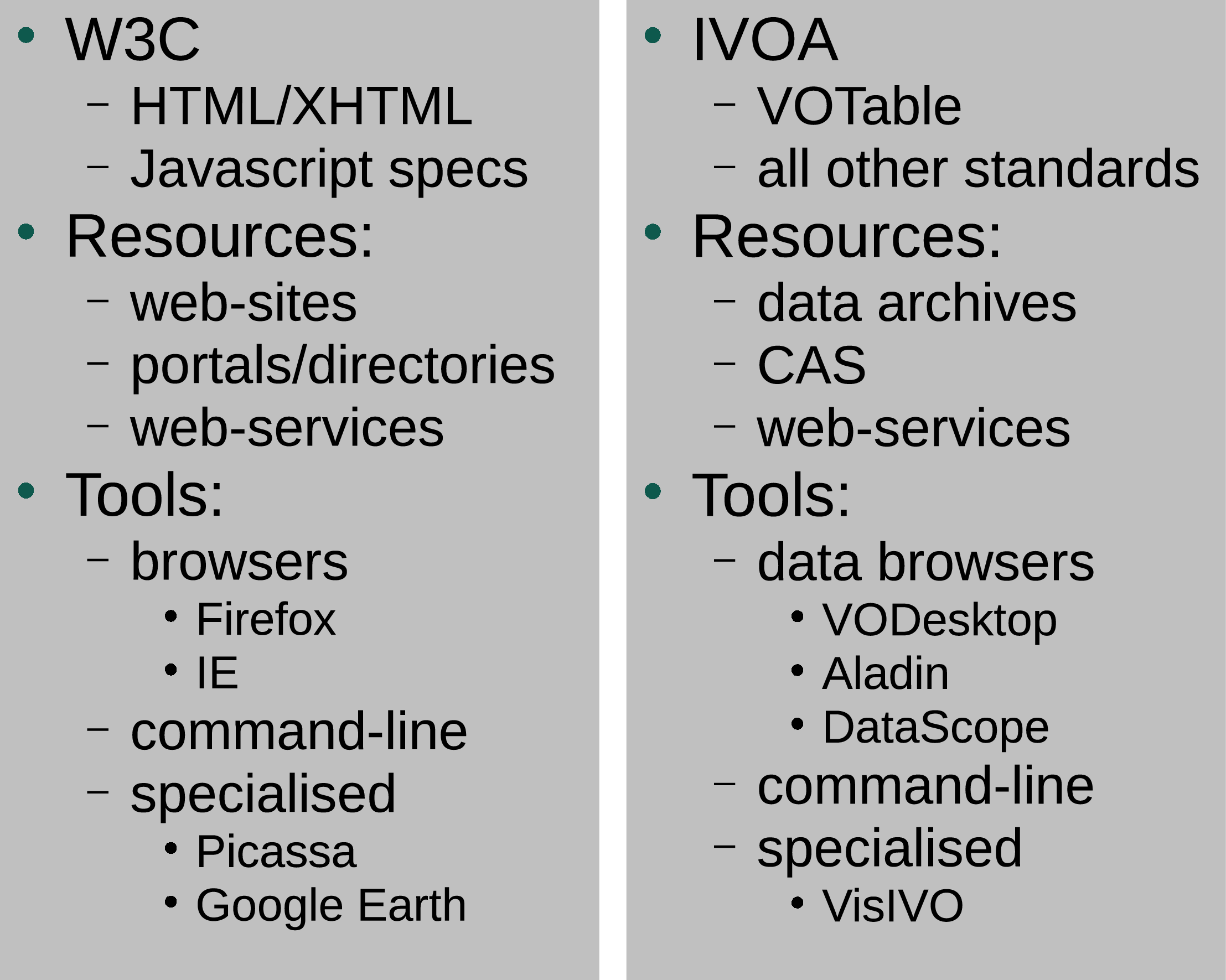}
\caption{Similarities between the concepts of WWW (left) and VO (right).\label{figvowww}}
\end{figure}

Virtual Observatory is sometimes referred as a World Wide Web for
astronomers. Indeed, there are numerous remarkable similarities between the
concepts of WWW and VO (Fig~\ref{figvowww}):
\begin{itemize}
\item \emph{IVOA} plays a similar role for the VO as \emph{W3C} does for WWW: these are
administrative organisms responsible for the definition of interoperability
standards. As examples, we can consider specifications of HTML/XHTML
developed by W3C and VOTable developed by IVOA
\item \emph{Resources} are the inalienable parts of both concepts. In case of WWW
these comprise: (1) web-sites, (2) portals and directories, (3)
web-services. In the VO we have: (1) data archives, (2) catalogue access
services (e.g. SDSS CASJobs: http://cas.sdss.org/), (3) astronomy-oriented web-services
\item \emph{Tools} represent another cornerstone of both, VO and WWW: (1) in
WWW we deal with web-browsers (e.g. {\sc Firefox}, {\sc Internet Explorer},
{\sc Safari}) while in the VO we have data browser and discovery tools, such
as {\sc astrogrid vo desktop}, {\sc cds aladin}, {\sc nvo datascope}; (2)
advanced users often deal with command-line tools to access the resources
such as {\sc curl} or {\sc wget} for WWW, and, similarly, VO clients based
on access libraries such as {\sc astro-runtime}; (3) finally, there are
specialised clients using WWW/VO protocols as infrastructure and/or data
transport, for instance, {\sc picassa} or {\sc google earth} with their
``analogues'' in the VO world such as {\sc VisIVO} 
\end{itemize}

\section{VO Science and Technology Now}

In this section I would like to briefly mention the existing accomplishments
of the Virtual Observatory. 

On the side of IVOA we have a comprehensive set of standards including: data
formats (VOTable), VO resource description (Resource Metadata), data model
for 1D spectra (Spectrum Data Model) and much more complex and general
Characterisation Data Model, Astronomical Data Query Language, protocols to
access images and spectra, application messaging protocol allowing different
VO tools to talk to each other, authorisation and authentication mechanisms,
and others. Many more standards are still at different phases of
development. Now it became possible to handle even very complex astronomical
datasets in the Virtual Observatory, such as 3D spectroscopy
\citep{CBLM06,ChilVO08} and results of N-body simulations.

In the meantime, application developers have created an impressive set of
VO-enabled tools from those of general interest to very specialised
applications. Many of them were presented in the review by M.~Allen (this
conference).

Data and service providers have contributed to the Virtual Observatory by
providing access to numerous data collections and archives at wavelength
domains from gamma-ray to radio. First services to access theoretical models
(e.g. theoretical spectra of stellar atmospheres in Spanish-VO or {\sc
pegase.2 / pegase.hr} synthetic stellar populations \citep{FR97,LeBorgne+04}
in VO-France, access to the results of cosmological simulations in Italian
VO) started to appear recently. We should also mention first prototypes of
data analysis services and value-added services associated with data access
services, such as modelling of the spectrophotometric properties of merging
galaxies in the GalMer database \citep{DiMatteo+08}.

The Virtual Observatory has been used for astronomical research for almost 5
years. The first VO-science result was the discovery of optically faint
obscured quasars by \citet{PARW04}. This was an example of a
multi-wavelength study carried out entirely inside the VO infrastructure.
Three years later, the VO studies of obscured AGN were continued
\citep{Richards+07}. 

A number of refereed papers were published by the Spanish VO project
members, presenting discoveries of unique objects done with the VO tools
\citep{CS07,CS08,CD08}.

A paper by \citet{Bayo+08} presenting VO SED Analyzer is of particular
interest, as the first refereed paper presenting a ``virtual instrument'', i.e.
a service in the VO aimed at data analysis, as well as its application to a
particular research project.

Many other studies made use of the VO tools and infrastructure combining
them with proprietary data access and analysis. For example, in
\citet{CM08}, authors used the VO data discovery and access mechanisms to
collect all existing data on a newly discovered object. This example
demonstrates how difficult may be to define the concept of a VO study or
VO-enabled study.

Nearly all research projects mentioned above used Virtual Observatory to do
data discovery, data access, and data mining. Therefore, the principal
question I would like to address is: Is it already possible to go beyond
data mining? The answer is: Yes, it is.

\section{VO Science Beyond Data Mining}

In this section I describe two VO-enabled research projects making heavy use
of VO technologies beyond data mining. The VO is used in connection with
dedicated observations and numerical simulations demonstrating the
proof-of-concept for such complex studies.

\subsection{Compact Elliptical and Tidally Stripped Galaxies}

This study was inspired by the serendipitous discovery of a very rare
compact elliptical (cE) galaxy in the central part of the nearby galaxy
cluster Abell~496 \citep{Chilingarian+07} which became the 6th known galaxy
of this class in the Universe. 

Compact elliptical galaxies have similar luminosities ($-18 < M_B <
-15$~mag) and stellar masses to dwarf ellipticals, but 10 smaller effective
radii ($r_e \sim 100 - 200$~pc) and, therefore, 100 times higher surface
brightness and 1000 times larger stellar density per unit of volume. The
prototype of the cE type is Messier~32, a satellite of the Andromeda galaxy.
All known cEs reside in the vicinities of larger stellar systems and/or in
the innermost regions of galaxy clusters.

Compact ellipticals are thought to be tidally stripped
intermediate-luminosity galaxies \citep{NP87, BCDG01, Graham02}, additional
arguments for this scenario have been provided by \citet{Chilingarian+07}
from stellar populations. However, low statistics did not allow us to
uniquely argue for this scenario of cE formation. 

Given small spatial sizes of cEs they become spatially-unresolved 
in ground-based observations for distances beyond $\sim$50~Mpc. Their
broadband optical colours well resemble K-type Galactic stars giving them
little chances to be included into the samples of large spectroscopic
surveys such as SDSS.

The key advantage here is provided by the Hubble Space Telescope, which can
efficiently resolve these little galaxies up-to a distance of 200~Mpc,
allowing us to study their structural properties. We realised this in the
course of our study of the Abell~496 cluster of galaxies
\citep{Chilingarian+08}, and decided to search for cE galaxies using the
power of the Virtual Observatory to study the role of tidal stripping in the
galaxy evolution. All details about this project will be soon provided in
Chilingarian et al. (in prep.), here we give a brief overview.

We have constructed a workflow including the following steps:
\begin{enumerate}
\item querying Vizier Catalogue Service to retrieve a list of galaxy
clusters having $z<0.055$;
\item querying NED to retrieve their precise coordinates and values of
Galactic extinction;
\item querying fully-reduced
direct images obtained with HST WFPC2 and ACS from the Hubble Legacy Archive
(HLA) using Simple Image Access Protocol (SIAP); 
\item running SExtractor \citep{BA96} as a remote tool on these images (no image download is required); 
\item selecting extended objects
having low ellipticity, effective radii below 0.7 kpc and $B$-band mean
effective surface brightness higher than 20 mag/arcsec$^2$; 
\item querying NED to check if there are published redshifts for the selected objects
and obtaining additional photometric data
\end{enumerate}

Having applied the workflow to the entire WFPC2 data collection in the HLA
we ended up with the archival images of 63 clusters with several dozens
candidate cE and tidally stripped galaxies in 30 of them. We found a large
number of objects in the scarcely populated region of the
$R_{e}$~vs~$\langle \mu \rangle_{e}$ and $M$~vs~$\langle \mu \rangle_{e}$
diagrams reflecting structural properties of galaxies.

\begin{figure}
\includegraphics[width=\hsize]{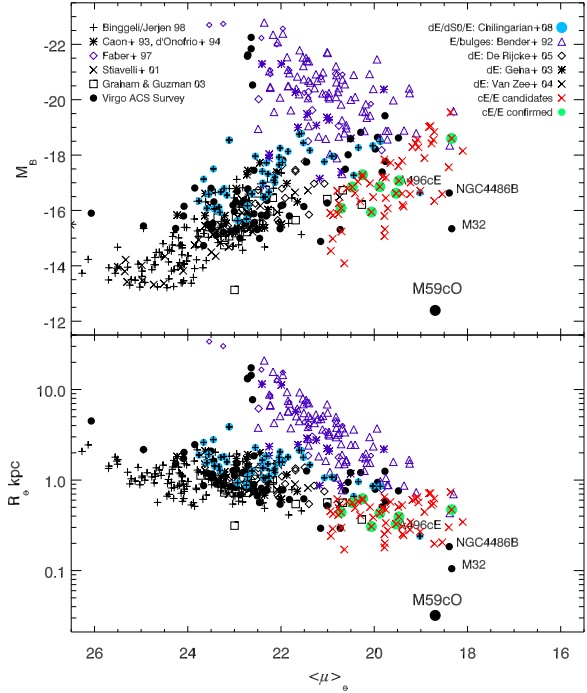}
\caption{Structural properties of early-type galaxies. See sources of data
in the text. New cE and tidally stripped galaxy candidates discovered using
the VO are shown by red crosses on-top of green circles for objects with
spectroscopic confirmation.\label{figMBmuav}}
\end{figure}

In Fig~\ref{figMBmuav} we present the structural properties of newly
discovered cE and tidally stripped galaxy candidates in comparison with
dwarf
\citep{BJ98,Stiavelli+01,GG03,GGvdM03,vZBS04,deRijcke+05,Ferrarese+06,Chilingarian+08},
intermediate-luminosity \citep{BBF92}, giant
\citep{BBF92,CCD93,DCC94,Faber+97} early-type galaxies, three nearby compact
ellipticals and transitional cE/UCD object \citep{CM08}.

Our workflow may have confused cE galaxies with (1) foreground or cluster
compact star-forming galaxies; (2) background giant early-type galaxies
hosting bright active nuclei (AGN); (3) background post-starburst galaxies.
The star-forming galaxies can be discriminated automatically by their blue
colours if multi-band data are available or manually by clumpy morphology.
The two remaining cases may arise when the distance to the background object
is 2--3 times the distance to the cluster, i.e. $<$400--500~Mpc ($z<0.12$)
in our study. Then, AGNs can be ruled out by checking X-ray point source
catalogues, and the probability of having a post-starburst galaxy at this
redshift is very low \citep{Goto07}.

The next stage of the project was to obtain high-quality optical
spectroscopic data on some of the candidates.

\begin{figure*}
\includegraphics[width=0.4\hsize]{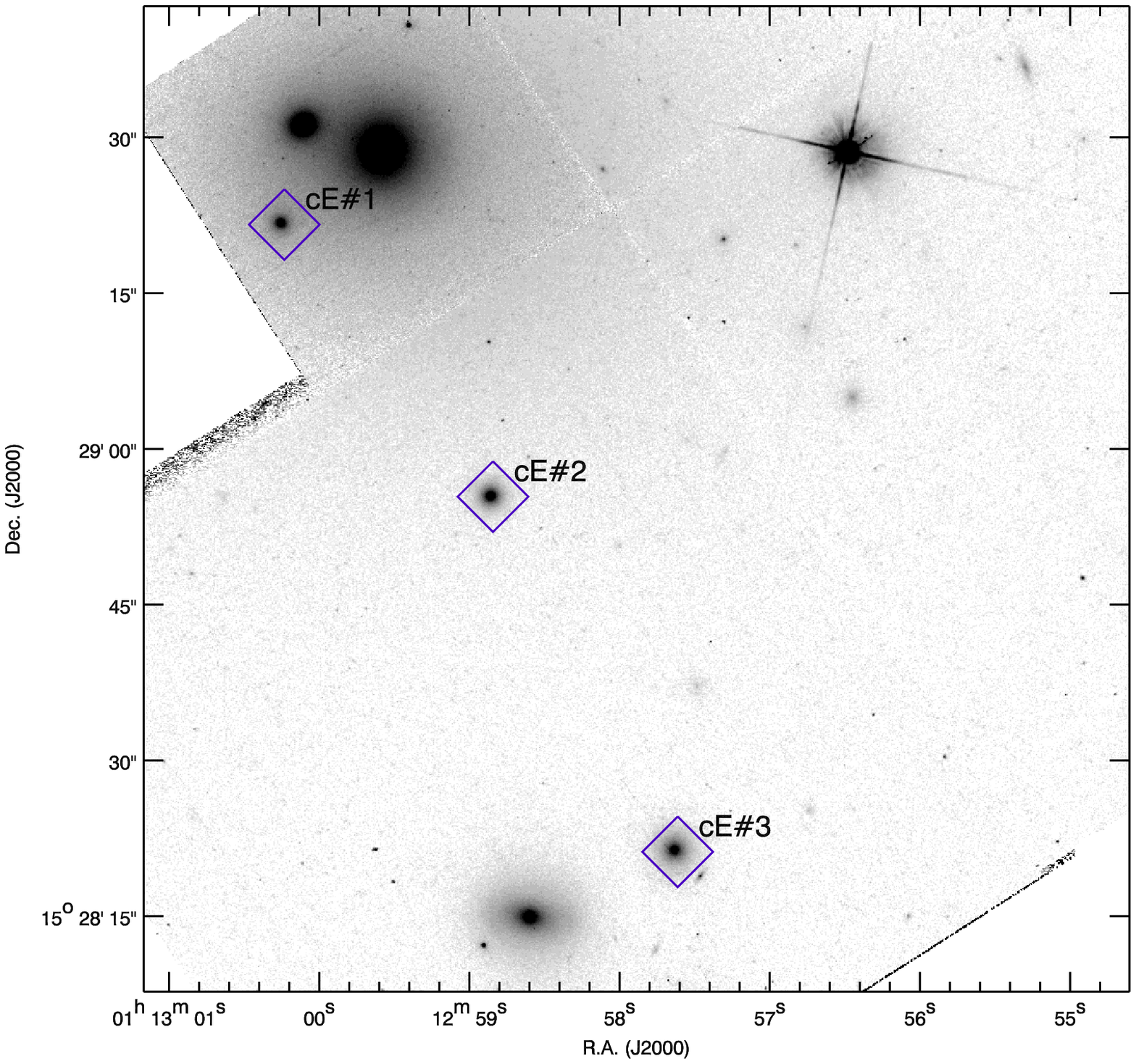}
\includegraphics[width=0.6\hsize]{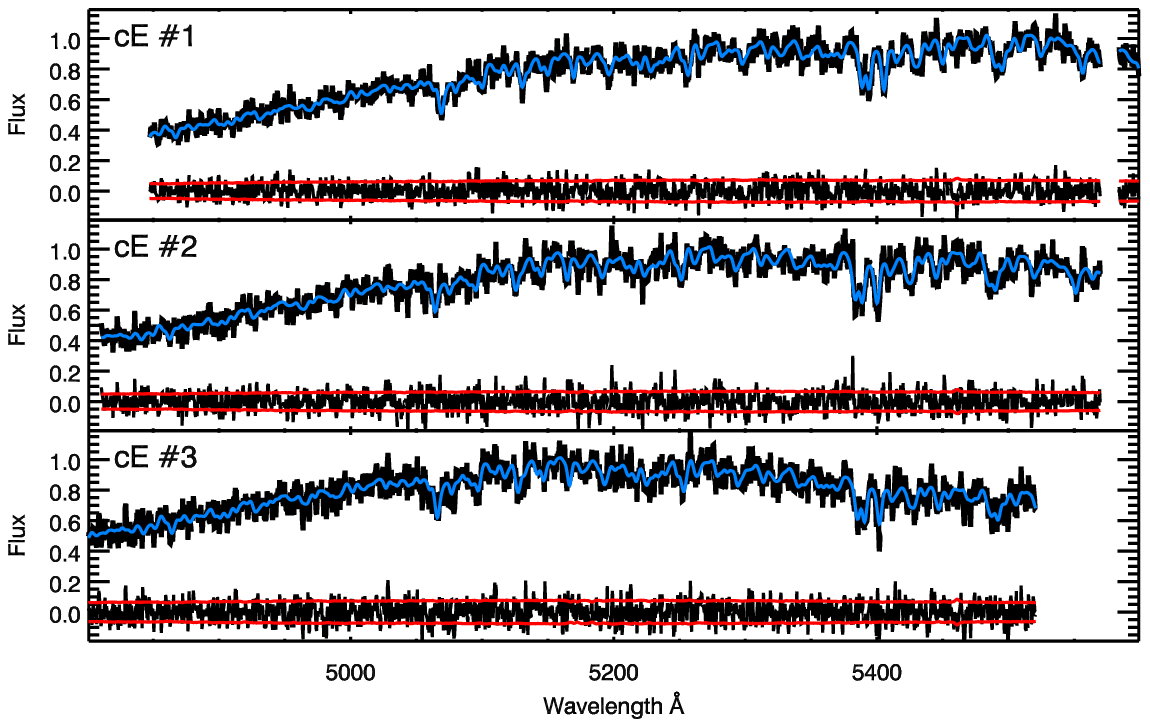}
\caption{HST WFPC2 image of the Abell~160 cluster with three confirmed cE
galaxies outlined (left). Spectra of these three objects obtained with the
SCORPIO universal spectrograph (right) with their best-fitting templates as
provided by the {\sc nbursts} full spectral fitting technique.\label{figA160}}
\end{figure*}

We have observed three galaxy clusters, Abell~160 (Fig~\ref{figA160}),
Abell~189, and Abell~397 hosting 8 candidate galaxies ($-17.0 <M_{I}<
-19.5$~mag) with the multi-slit unit of the SCORPIO spectrograph
\citep{AM05} at the Russian 6-m ``Bol'shoi Telescop Azimutal'nyy'' (BTA) in
August 2008. We have analysed the spectra by fitting them against
high-resolution {\sc pegase.hr} stellar population
models using the novel {\sc nbursts} full spectral fitting technique
\citep{CPSK07,CPSA07}, and obtained precise radial velocities, internal
velocity dispersions, luminosity-weighted stellar ages and metallicities.
All 8 candidate objects were confirmed to be cluster members having stellar
populations older than 8~Gyr with metal abundances typically between
$Z_{\odot}/2.5$ and the solar value ($Z_{\odot}/5$ for one object) and
internal velocity dispersions between 50 and 100~km~s$^{-1}$.

We have performed numerical simulations of tidal stripping of
intermediate-luminosity early-type disc galaxies by the potential of the
galaxy cluster including the central cD galaxy using the {\sc gadget-2} code
\citep{Springel05}. The simulations suggest that the progenitor galaxies may entirely
lose their discy components due to tidal stripping, while keeping the bulges
although with significant stellar mass loss as well. The remnants of the
tidal stripping for some initial conditions and orbital configurations well
resemble the observed properties of cEs, although in most cases they are
still remain quite extended. This explains why cE galaxies are not very
common.

We have faced a number of issues while undertaking this project, most of
them are infrastructural.
\begin{itemize}
\item There is no real VO access to NED. We had to develop and use
``home-made'' scripts to execute queries.
\item SIAP interface in the HLA is not publicly announced, although it is
used internally in the project. The service URL has been provided privately
to us by the HLA developers, although it was possible to get it by reverse
engineering of the JavaScript code.
\item We had to setup a customized SExtractor service and spend a lot of
time to fine-tune it by using some undocumented features.
\item It took 3 semesters to convince the TAC to approve our telescope
proposal.
\item We were the first to use the SCORPIO multi-slit mode with the
high-resolution grism, therefore it was necessary to design and develop the
data reduction pipeline.
\end{itemize}

The main result of our study is that the class of cE galaxies is converted
from ``unique'' into ``common under certain environmental conditions''. We
provide evidences for the importance of tidal stripping of stellar discs as
a way to create peculiar early-type galaxy population in the cluster
centres. Now we can explain the existence of very strange objects such as
VCC~1199 having supersolar metallicity for a very low luminosity of $M_B
\approx -15$~mag.

This was the first study, where the primary step was done in the VO, then
the discovered objects were followed-up with a large telescope and
successfully reproduced by numerical simulations.

\subsection{Optical and Near-Infrared Galaxy Colours}

While preparing my presentation for this meeting, I decided to make
something special. The challenge was to get valuable results related to
studies of galaxies \emph{in one week} using the VO starting the project
from scratch. My main collaborator in this project was I.~Zolotukhin,
located geographically in a different place, so we had to work remotely and
interact only online.

We decided to study optical and near-infrared colours of nearby galaxies and
try to connect them to their stellar population properties. NIR magnitudes
are less sensitive to the stellar population age compared to the optical
colours, therefore they can be used as better stellar mass tracers (although
not perfect). In addition, the effects of extinction inside the galaxies
being studied are less important in the NIR spectral bands. Spectroscopic
information on stellar ages and metallicities should become additional
important bricks of information. The catalogue will be presented in
Chilingarian, Zolotukhin \& Melchior (in prep.)

We have used the following resources:
\begin{itemize}
\item SDSS DR7 \citep{Abazajian+08} photometric catalogues as a source of optical magnitudes
\item SDSS DR7 spectra to get stellar population properties by the full
spectral fitting
\item UKIDSS \citep{Lawrence+07} DR4 Large Area Survey (LAS) catalogue as a source of NIR magnitudes
\end{itemize}

The techniques we have exploited:
\begin{itemize}
\item position-based cross-match (possible to do in the VO)
\item stellar population modelling using {\sc pegase.2}/{\sc pegase.hr}
(possible to do in the VO)
\item {\sc nbursts} full spectral fitting technique (yet as a stand-alone non-VO
service)
\end{itemize}

From the zoo of VO tools we selected {\sc topcat/stilts} to join and merge
large tables, scripts-based access to SDSS, and {\sc astrogrid vo desktop}
to access UKIDSS catalogues and perform the cross-match.

We have used SDSS CASJobs to select all galaxies from the spectroscopic
sample having redshifts $0.03 < z < 0.3$ in the SDSS stripes 9 to 16 which
have been partially covered by the UKIDSS. This query has returned
approximately 170 thousand objects. 

Then, we have cross-matched this list against the UKIDSS DR4 LAS catalogue
with a search radius of 5~arcsec. This step can be done either using WFCAM
Archive or through the {\sc astrogrid vo desktop} application using the
multi-cone search interface. Notice, that the access to UKIDSS DR4 was
restricted, therefore to query it in an automatic way it was necessary to
use the authorisation/authentication mechanisms provided by the VO.

Then we have computed and applied $K$-corrections by fitting optical-NIR
spectral energy distributions (SEDs) against {\sc pegase.2} stellar
population models to get rest-frame magnitudes.

At the final step, we have processed all selected SDSS DR7 spectra using the
{\sc nbursts} full spectral fitting technique in order to estimate velocity
dispersions, ages, and metallicities of all galaxies in 3-arcsec wide
apertures.

The non-trivial problem is the homogenisation of the photometric data.
Firstly, we tried to use the Petrosian magnitudes provided in both
catalogues, but we quickly realised that due to very different sensitivity
of the two surveys the Petrosian radii used to measure the magnitudes may be
very different and the difference is correlated with the galaxy colours.
Finally, we decided to deal with fluxes in the 3~arcsec wide apertures,
which are provided directly by SDSS and easy to compute for UKIDSS using the
three provided aperture magnitudes. These magnitudes may not reflect the
real total colours of galaxies, since our fixed aperture corresponds to
different spatial sizes at different redshifts, but, for explaining the
colour properties using additional spectroscopic information obtained by the
SDSS in the same apertures, this approach is preferable.

Another important problem is the $K$-correction, or dependence of colours on
the redshift due to the fact that the filter transmission curves effectively
contain different regions of galaxies' SEDs at different redshifts. There
are several existing prescriptions for the computation of $K$-corrections
\citep{Mannucci+01,BR07}, however, they provide controversial information
for the NIR spectral bands. Therefore, we have decided to deduce
$K$-corrections from the multi-wavelength SED fitting of the data against
the {\sc pegase.2} SSP models redshifted according to the spectroscopic
information and varying the effects of internal dust extinction. The
behaviour of $K$-corrections in the optical bandpasses obtained in this
fashion well resembles the results of \citet{BR07}. At the same
time, our NIR $K$-corrections turned to be very different, however, well
corresponding to those presented in \citet{Mannucci+01}. 

Since we had optical spectra available from SDSS for all the galaxies in our
sample, we computed the actual values of flux differences by integrating the
spectra in the rest-frame and redshifted filter bandpasses unless they moved
out of the spectral coverage. Given the wavelength range of SDSS spectra we
were able to compute the ``true'' $K$-corrections for the $g$ band for any
redshift, for the $r$ band for objects having $z<0.25$, and for the $i$ band
till $z<0.08$. These ``true'' values perfectly agree with the values derived
from our SED fitting with a typical errors not exceeding 0.05~mag.
Therefore, we conclude that our prescriptions for the computation of
$K$-corrections have reasonable quality for the studies of optical/NIR
galaxy colours. The full discussion related to the computation and
applications of $K$-corrections for nearby galaxies will be provided in
Chilingarian, Melchior \& Zolotukhin (in prep.)

We have also faced a number of technical issues:
\begin{itemize}
\item SDSS DR7 is not accessible via VO protocols, therefore we used its own
CASJobs portal.
\item There were numerous problems accessing UKIDSS DR4 through the VO due
to bugs in the implementation of services and access interfaces. However,
all these questions have been solved very efficiently by the UKIDSS and
Astrogrid teams.
\item Need to download, upload, merge, and convert lengthy tables.
\item Still there is no way to perform the cross-match against user-uploaded
table using the ADQL queries: these mechanisms are still to be implemented
\end{itemize}

\begin{figure}
\includegraphics[width=\hsize]{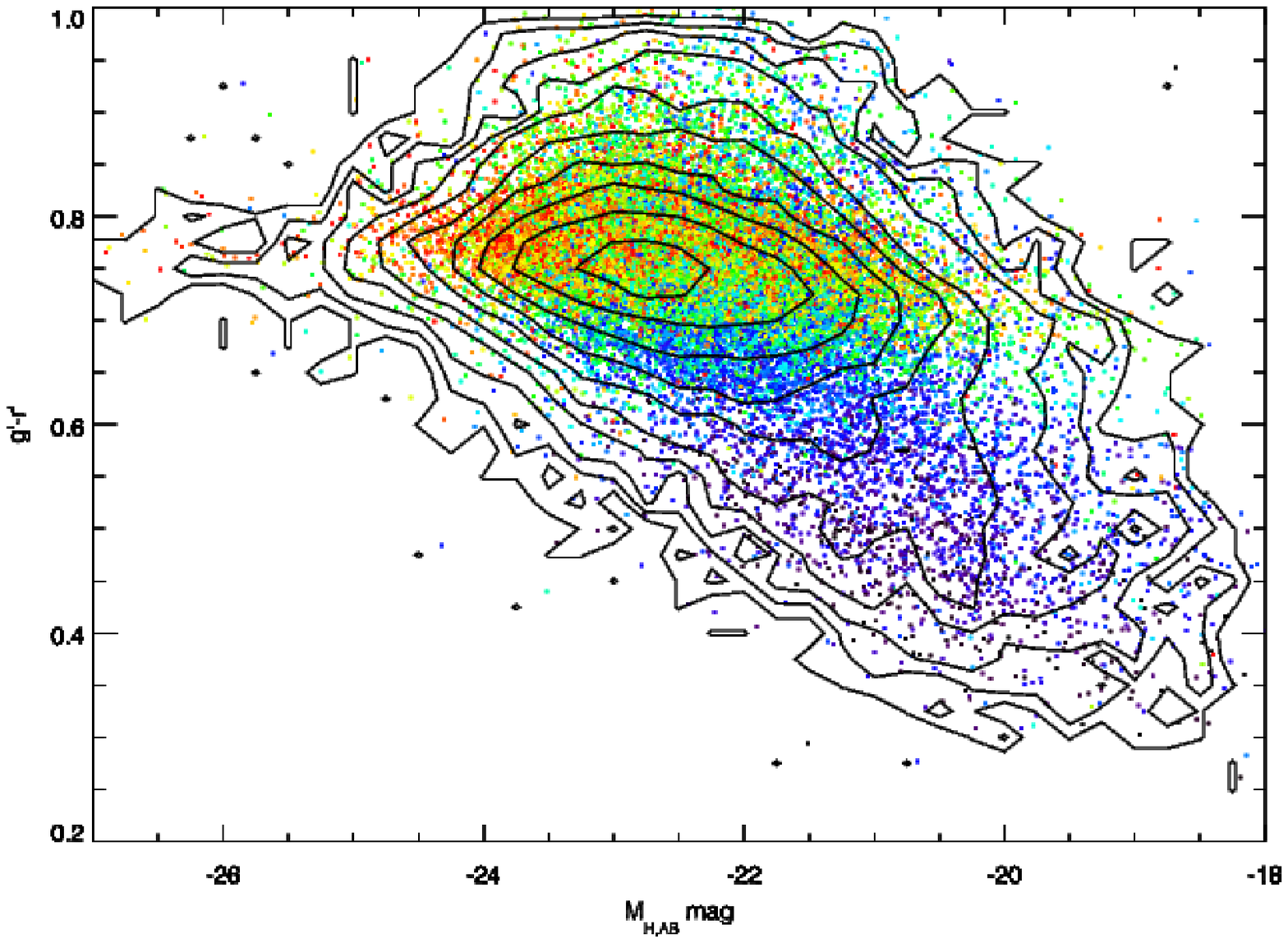}\\
\includegraphics[width=\hsize]{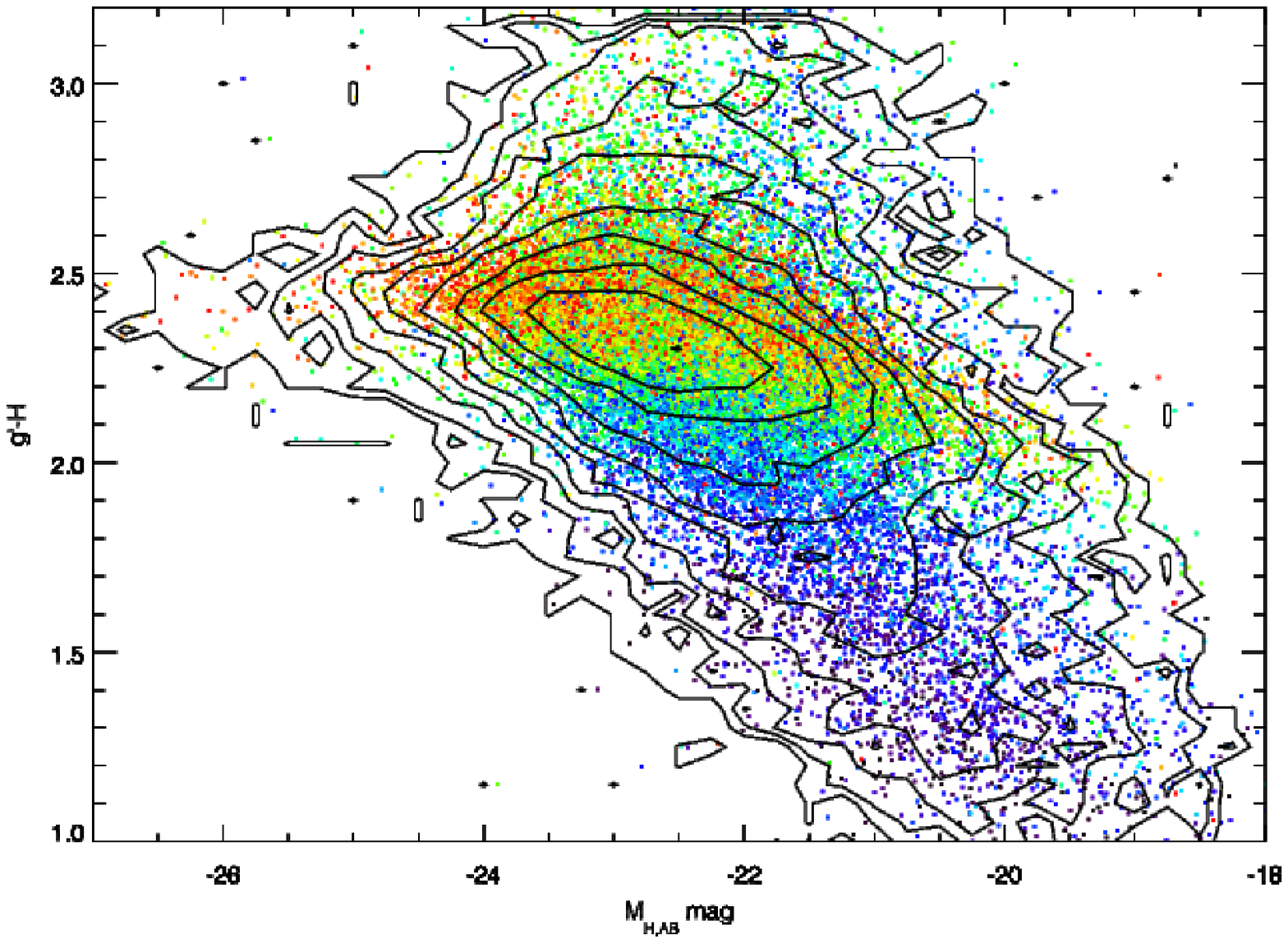}
\caption{Colour-magnitude diagrams for nearby galaxies: optical and NIR
colours. All magnitudes are extinction- and $K$-corrected. The top and
bottom panels present $g - r$ vs $M_H$ and $g - H$ vs $M_H$ respectively.
The SSP-equivalent ages are colour coded, values from violet to red
correspond to 1 to 15~Gyr.\label{figRS}}
\end{figure}

In Fig~\ref{figRS} we present the colour--magnitude plots for the galaxies
from our sample. We use the rest-frame $H$ magnitude which is much less
sensitive to the effects of stellar age than optical bands. Spectroscopic
ages obtained from the full spectral fitting are colour-coded. It is
immediately evident that the red sequence \citep{Strateva+01} is populated
by old galaxies, while in the blue cloud there is a significant age
gradient.

There is a 3-magnitude long high-luminosity tail of the red sequence clearly
seen on the plots, providing an immediate evidence that such galaxies cannot
be formed by equal-mass mergers of objects from the blue cloud. 

Another feature seen in the $g - H$ vs $M_H$ plot is a population of young
and intermediate-age objects overlapping the red sequence and sometimes
being as much as 0.7~mag redder in the $g - H$ colour. These are probably
dusty galaxies with active star formation, so the superposition of young and
all populations, plus dust attenuation creates such an appearance.

This project is still very far from being finished. We foresee to add GALEX
ultraviolet data and fit SDSS spectra together with photometric data points
with more realistic galaxy models than simple SSPs, for example, to include
two star formation episodes with different dust attenuation for them. We
also plan to study the distribution of emission line strengths since we are
able to precisely model the stellar population, making feasible studies of
even very faint emission lines.

\section{Summary}

The Virtual Observatory is already at the production level. Scientists not
associated directly to VO projects are trying to feel the way. The
scientific results already obtained are impressive and very important as a
proof-of-concept. The advantages of the VO approach are clear: one can
transparently access and process enormous volumes of data from different
sources. But, of course, the VO should not be considered as replacement
for scientists -- it is just a tool to help them.

In my opinion, the major problem for a scientist in the VO now is very
little, but numerous and, therefore, annoying infrastructural faults: all the
individual bricks exist, but them together still requires a lot of efforts.

\section*{Acknowledgments}

Author is grateful to the organizing committee of the workshop for
invitation, to collaborators in the projects presented in this article,
especially to Ivan Zolotukhin, Veronique Cayatte, and Anne-Laure Melchior,
to ESO for supporting financially the attendance to the workshop. The
research presented here is partially based on the data obtained with the
Hubble Space Telescope and by the Sloan Digital Sky Survey project. The SDSS
web-site is http://www.sdss.org/

\end{document}